# Reduced Order Fractional Fourier Transform — A New Variant to Fractional Signal Processing: Definition and Properties

*Sanjay Kumar*

*Department of ECE, Thapar Institute of Engineering and Technology, Patiala, Punjab, India*

E-mail: *sanjay.kumar@thapar.edu*

____________________________________________________________________________

**Abstract**

In this paper, a new variant to fractional signal processing is proposed known as the *Reduced Order Fractional Fourier Transform (ROFrFT)*. Various properties satisfied by its transformation kernel is derived. The properties associated with the proposed ROFrFT like shift, modulation, time-frequency shift property are also derived and it is shown mathematically that when the rotation angle of ROFrFT approaches $\pi/2$, the proposed ROFrFT reduces to the conventional Fourier transform (FT). Also, the ROFrFTs of various kinds of signals is also derived and it is shown that the obtained analytical expressions of different ROFrFTs are a reduced form of the conventional fractional Fourier transform (FrFT). It is also shown that proposed definition of FrFT is easier to be handled analytically. Finally, the convolution theorem associated with the proposed ROFrFT is derived with its various properties like shift convolution, modulation convolution, and time-frequency shift modulation properties. It has been shown that with this proposed new definition of FrFT, the convolution theorem has been reduced to multiplication in the fractional frequency domain — an exciting result, same as that of Euclidean Fourier transform.

# 1. Introduction

The characteristics of signals of interest determines the selection of signal processing techniques to be applied. Various frequency-based and time-frequency (TF) based techniques are applied and mentioned in popular literature [1] [2] [3].

In the conventional signal processing methods, the two natural existing variables namely, time ($t$) and frequency ($f$) are used exclusively and independently of each other, whereas in the TF signal processing methods, these variables are used *concurrently* [3]. This means that rather than viewing a one-dimensional signal and some transform, the TF signal analysis and processing investigates a two-dimensional signal, which is obtained from the signal through some TF transform.

The conventional frequency-based signal processing methods rely on Fourier transformation, whereas the TF signal analysis and processing involves many TF transforms, namely short-time Fourier transform (STFT), wavelet transform (WT), bilinear TF distribution/Wigner distribution (WD), Gabor-Wigner distribution (GWD), fractional Fourier transform (FrFT), linear canonical transform (LCT), Stockwell transform (ST) and to name a few [3].

The ubiquity of the original Fourier transform (FT) has proliferated many important signal and image analysis applications. It is known as a powerful tool which reveals the overall spectral contents by assuming that the given signal is stationary in nature. However, this assumption of stationarity is an ideal assumption and not particularly useful in practical applications [4], since most practical signals of interest are non-stationary in nature, which have dynamic frequency content. Thus, it is necessary to resort to the TF representation (TFR) that represents the energy density of a signal simultaneously in the TF plane.

Thus, the TF transforms are of great interest among signal processing researchers in varied allied real-time applications due to their natural decomposition of a signal into a function which is localized in both time and frequency. An excellent survey on the state of art of the TF signal analysis and processing and its applications was given in [1] [3] [5] [6].

In the recent years, along with the various TFRs of signals, various researchers are using the concept of fractional Fourier transform (FrFT), which extends the capabilities of the conventional FT. The FrFT is a generalization of the conventional FT, which depends on an additional parameter and can be interpreted as a rotation in the TF plane [7] [8]. The FrFT has been extensively applied for significant applications in digital signal [9] [10] [11] [12], image processing [13] [14] [15] [16] as well as in quantum mechanics and optical information processing [17] [18] [19].

It is well-known that the conventional FT can be used to adequately represent signals in terms of the sinusoids in the frequency domain, and applying the FrFT allows to analyze a given signal in the TF domain and thus, it can be used for signals having non-sinusoidal basis such as linear chirp signals [20]. The advantages of using FrFT-based methods includes high resistance to noise as well as low computational complexity [19] [20], its practical applications are expected to grow significantly in years to come, given that the FrFT offers many advantages over the conventional Fourier analysis, including signal/image restoration and noise removal [34].

In this paper, a mathematical investigation is done to determine a new and novel definition of the FrFT — the *Reduced Order Fractional Fourier Transform* (ROFrFT), which could prove beneficial to fractional signal processing applications and it has not been involved in any literature so far. Its definition with various properties associated with it and its transformation kernel have been derived analytically. Next, thorough analytical derivations is done to determine various analytical expressions of different kinds of signals which depends on the rotation angle. Finally, the convolution theorem associated it is derived with detailed derivations of shift, modulation and time-frequency shift properties.

The rest of the paper is organized as follows. In Section 2, the preliminaries of the reduced order fractional Fourier transform (ROFrFT) is presented, with various properties of its transformation kernel in Section 3 and various properties associated with ROFrFT in Section 4 are investigated along with their analytical proofs. In Section 5, a mathematical investigation is presented for deriving ROFrFTs of different kinds of signals. The new definition of convolution theorem of ROFrFT which will prove to be much beneficial for fractional signal processing society is presented in Section 6 and finally conclusions and the future scope of the proposed work for the fractional signal processing society is summarized in Section 7.

**2. Reduced Order Fractional Fourier Transform: Definition and Integral Representation**

As it well known that the fractional Fourier transform (FrFT) is a generalization of the conventional Fourier transform (FT), which was introduced from analytical aspect by Namias and appeared in many applications in optics field at that time. It was around 1990s that its potential capability has appeared in signal processing society. However, with the advancement of science and technology domains, the FT has faced many shortcomings in its nature, where its orthogonal basis are sinusoidal in nature.

For processing practical signals, such as biomedical signals, seismic data, radar, sonar, audio, video signals, etc. the Fourier transformation is not capable to extract the useful information from them and that too in non-stationary environment, so there's a need of signal processing transformation which could process non-stationary signals in non-stationary environment faithfully. Many rich literature are available where various signal processing researchers apply FrFT to solve daily real life problems, whether they apply it for biomedical, seismic, wireless, radar, sonar, audio, video processing etc. applications.

The added advantage of using FrFT tool lies in its nature ─ the FrFT is a linear transformation where its orthogonal basis is a linear frequency modulation (chirp) signal, so it has the capability to process non-stationary signals efficiently as compared to the conventional FT. Also, it is more flexible and suitable for processing chirp-like signals due to the additional degree of freedom as compared to FT [8], [9], [11], [17], [18], [19]. Additionally, through the usage of FT, the whole spectrum of the signal of interest is obtained and it cannot obtain the local TF characteristics of the signal, which is essential for non-stationary signal processing. So many novel signal processing transforms have been discovered by researchers, which has their own advantages and limitations [21].

However, it's been a decade that the research on FrFT has proliferated in varied research areas of communication, radar, sonar, image, etc. The FrFT can be interpreted as a rotation of the signal in the TF plane by an angle [22]. The time-domain and frequency-domain representations are the two special cases of the FrFT. A fundamental advantage of using FrFT for signal filtering is that the signal of interest can be represented in any domain within the range of the rotation angle, rather than being limited to only either in the time-domain or in the frequency-domain [22]. Mathematically, the FrFT implements the order parameter $\varphi$, which acts on the conventional FT operator. To say, the $\varphi^{th}$ order FrFT represents the $\varphi^{th}$ power of th conventional FT operator [8].

The FrFT is a linear operator with a fractional Fourier order parameter $a$ or transform parameter $\varphi$, which corresponds to the $a$th fractional power of FT operator, $\mathcal{F}$, and can be viewed as a counterclockwise rotation by a fractional order $\varphi$ in the TF plane. The FrFT of a signal $x(t)$ is defined as [9], [17], [18], [19]

$$\mathbb{F}_{\mathcal{F}}^{\varphi}(x(t)) = X^{\varphi}(u_{\varphi}) = \int_{-\infty}^{\infty} x(t) K_{\varphi}(t, u_{\varphi}) dt \qquad (1)$$

where $0 < |a| < 2$, the transformation kernel, $K_{\varphi}(t, u_{\varphi}) = \sqrt{\frac{1 - j \cot \varphi}{2\pi}} e^{j\left(\frac{t^2 + u_{\varphi}^2}{2}\right) \cot \varphi - j u_{\varphi} t \csc \varphi}$ with the transform angle $\varphi = a\pi/2$ [17], $j^2 = -1$ and $\mathbb{F}_{\mathcal{F}}^{\varphi}$ denotes the FrFT operator. The $a$th FrFT domain makes an angle $\varphi = a\pi/2$ with the time domain in the time-frequency plane [12]. When $\varphi = \pi/2$, it converges to the classical FT and when $\varphi = 0$, it will be an identity operation. It is seen that the FrFT transforms a signal into an intermediate domain between time and frequency when $\varphi \neq N(\pi/2)$, where $N$ is an integer [23]. Also, it can be inferred from (1) that the kernel $K_{\varphi}(t, u_{\varphi})$ is composed of chirp basis functions with a sweep rate of $\cot \varphi$. So the FrFT can be interpreted as signal decomposition into chirp functions. Thus, due to the property of concentration of LFM energy and its multi-domain nature, the FrFT has achieved a wider acceptance in the DSP community [24].

Based upon (1), it can be seen that the conventional FrFT can be realized in a four step process as mentioned in [9] [25]. As it was discussed in [26], for the fractional correlation operation with the optical implementation in the conventional FrFT domain, the chirp term $\exp\left(\frac{j}{2}u_\varphi{}^2 \cot\varphi\right)$ in the conventional FrFT definition (1) can be removed and one gets a reduced form of the FrFT definition that could prove to be beneficial for the optical information processing society.

So motivated by the research potential of [26], the proposed work emphasize the use of new definition of FrFT which could reveal magnificent property and could be applied in various science and engineering applications. Based on a comparison with the conventional FrFT, it is found out that the proposed definition of FrFT — ROFrFT, is easier to be handled analytically, and also all the analytical expressions obtained for different properties, convolution theorem and its properties are much easier to obtain analytically.

The Reduced Order Fractional Fourier transform (ROFrFT) of the signal $x(t)$ is represented by

$$\mathbb{F}_R^\varphi[x(t)] = X_R^\varphi(u_\varphi) = \int_{-\infty}^{\infty} x(t)\, \mathrm{K}_\varphi^R(t, u_\varphi)\, dt \qquad (2)$$

where $\mathrm{K}_\varphi^R(t, u_\varphi) = \sqrt{1 - j\cot\varphi}\, \exp\left[\frac{j}{2} t^2 \cot\varphi - j\, t\, u_\varphi \csc\varphi\right]$ \qquad (3)

Here, $t$ and $u_\varphi$ can interchangeably represent time and frequency domains. The transform output lies between time and frequency domains, except for the special cases of $\varphi = 0$ and $\varphi = \pi/2$. Based upon (2), the ROFrFT can be realized in a three step process, which is illustrated in Fig. 1 as follows:

(i) pre-multiplication of the input signal by a linear chirp with the frequency modulation (FM) rate determined by the transform order;

(ii) computation of the scaled FT ($\mathcal{F}$) with a scaling factor of $\csc\varphi$;

(iii) post-multiplication by a complex amplitude factor.

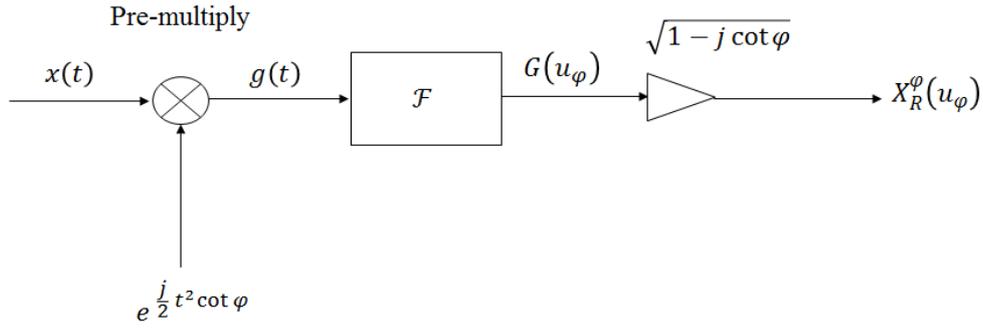

Fig. 1. ROFrFT block diagram.

## 3. Properties satisfied by the ROFrFT Transformation Kernel

The transformation kernel of ROFrFT is given by

$$K_\varphi^R(t, u_\varphi) = \sqrt{1 - j \cot \varphi} \, \exp\left[\frac{j}{2} t^2 \cot \varphi - j\, t\, u_\varphi \csc \varphi\right] \tag{4}$$

The transformation kernel of ROFrFT satisfies the following properties:

*(i) Diagonal symmetry*:

$$K_\varphi^R(t, u_\varphi) = K_\varphi^R(u_\varphi, t) \tag{5}$$

***Proof***: The right-hand side of (4) becomes $K_\varphi^R(u_\varphi, t) = \sqrt{1 - j \cot \varphi}\, \exp\left[\frac{j}{2} t^2 \cot \varphi - j\, u_\varphi\, t \csc \varphi\right]$. This shows the equivalence of (4). Hence proved.

*(ii) Complex conjugate*:

$$K_{-\varphi}^R(u_\varphi, t) = \overline{K_\varphi^R(u_\varphi, t)} \tag{6}$$

***Proof***: The left-hand side of (6) becomes

$$K_{-\varphi}^R(u_\varphi, t) = \sqrt{1 + j \cot \varphi}\, \exp\left[-\frac{j}{2} t^2 \cot \varphi + j\, u_\varphi\, t \csc \varphi\right]$$

and the right-hand side of (6) becomes

$$\overline{K_\varphi^R(u_\varphi, t)} = \sqrt{1 + j\cot\varphi}\, \exp\left[-\frac{j}{2}t^2 \cot\varphi + j\, u_\varphi\, t\, \csc\varphi\right]$$

Hence proved.

*(iii) Point symmetry*:

$$K_\varphi^R(-u_\varphi, t) = K_\varphi^R(u_\varphi, -t) \tag{7}$$

*Proof:* The left-hand side of (7) becomes

$$K_\varphi^R(-u_\varphi, t) = \sqrt{1 - j\cot\varphi}\, \exp\left[\frac{j}{2}t^2 \cot\varphi + j\, u_\varphi\, t\, \csc\varphi\right]$$

and the right-hand side of (7) becomes

$$K_\varphi^R(u_\varphi, -t) = \sqrt{1 - j\cot\varphi}\, \exp\left[\frac{j}{2}t^2 \cot\varphi + j\, u_\varphi\, t\, \csc\varphi\right]$$

Hence proved.

## 4. Properties associated with ROFrFT

This section derives the important analytical properties of the ROFrFT in detail. It is seen that the results are the generalizations of the basic properties of the FT.

**Theorem 4.1** (**Shift Property**): *Let $x(t) \in L^1(\mathbb{R})$. The ROFrFT of a shift by $\tau \in \mathbb{R}$ is given by*

$$\mathbb{T}_R^\varphi\{\mathbb{S}_\tau x\}(u_\varphi) = \mathbb{T}_R^\varphi\{x(t \pm \tau)\}(u_\varphi) = e^{\frac{j}{2}\tau^2 \cot\varphi \pm j u_\varphi \tau \csc\varphi}\, X_R^\varphi(u_\varphi \pm \tau\cos\varphi) \tag{8}$$

*Proof:* Replacing $x(t)$ by $x(t \pm \tau)$ in the integral representation (2) of ROFrFT, one gets

$$\mathbb{T}_R^\varphi\{x(t \pm \tau)\}(u_\varphi) = \sqrt{1 - j\cot\varphi}\, \int_{-\infty}^{\infty} x(t \pm \tau)\, e^{\frac{j}{2}t^2 \cot\varphi - j\, t\, u_\varphi \csc\varphi}\, dt \tag{9}$$

By assuming $(t \pm \tau) = t'$, i.e., $t = (t' \mp \tau)$ one solves (9) as

$$\mathbb{T}_R^\varphi\{x(t \pm \tau)\}(u_\varphi) = \sqrt{1 - j\cot\varphi}\, \int_{-\infty}^{\infty} x(t')\, e^{\frac{j}{2}(t' \mp \tau)^2 \cot\varphi - j(t' \mp \tau)\, u_\varphi \csc\varphi}\, dt' \tag{10}$$

Solving further, one gets

$$\mathbb{F}_R^\varphi\{x(t \pm \tau)\}(u_\varphi) = e^{\frac{j}{2}\tau^2 \cot\varphi \pm ju_\varphi \tau \csc\varphi} \sqrt{1-j\cot\varphi} \int_{-\infty}^{\infty} x(t') \, e^{\frac{j}{2}(t')^2 \cot\varphi - ju_\varphi(t' \pm \tau\cos\varphi)\csc\varphi} dt' \quad (11)$$

Thus, the ROFrFT of a shift by $\tau \in \mathbb{R}$ of a function $x(t \pm \tau)$ is given by

$$\mathbb{F}_R^\varphi\{\mathbb{S}_\tau x\}(u_\varphi) = \mathbb{F}_R^\varphi\{x(t \pm \tau)\}(u_\varphi) = e^{\frac{j}{2}\tau^2 \cot\varphi \pm ju_\varphi \tau \csc\varphi} X_R^\varphi(u_\varphi \pm \tau\cos\varphi) \quad (12)$$

If $\varphi$ equals $\pi/2$ in (12), the expression (12) reduces to $e^{\pm j\omega\tau} X(\omega)$, same as that of FT, where $\omega$ is the frequency variable in the FT domain.

**Theorem 4.2** (**Modulation Property**): *Let $x(t) \in L^1(\mathbb{R})$. The ROFrFT of a modulation by $q \in \mathbb{R}$ is given by*

$$\mathbb{F}_R^\varphi\{\mathbb{M}_q x\}(u_\varphi) = \mathbb{F}_R^\varphi\{e^{\pm jqt} x(t)\}(u_\varphi) = X_R^\varphi(u_\varphi \mp q\sin\varphi) \quad (13)$$

***Proof:*** For the frequency shifting, replace $x(t)$ by $\exp(\pm jqt)\, x(t)$ in the integral representation (2) of ROFrFT, one gets

$$\mathbb{F}_R^\varphi\{e^{\pm jqt} x(t)\}(u_\varphi) = \sqrt{1-j\cot\varphi} \int_{-\infty}^{\infty} \exp(\pm jqt)\, x(t)\, e^{\frac{j}{2}t^2 \cot\varphi - j t u_\varphi \csc\varphi} dt \quad (14)$$

Simplifying further,

$$\mathbb{F}_R^\varphi\{e^{\pm jqt} x(t)\}(u_\varphi) = \sqrt{1-j\cot\varphi} \int_{-\infty}^{\infty} x(t)\, e^{\frac{j}{2}t^2 \cot\varphi - ju_\varphi t \csc\varphi \pm jqt} dt \quad (15)$$

Solving further, one gets

$$\mathbb{F}_R^\varphi\{e^{\pm jqt} x(t)\}(u_\varphi) = \sqrt{1-j\cot\varphi} \int_{-\infty}^{\infty} x(t)\, e^{\frac{j}{2}t^2 \cot\varphi - j(u_\varphi \mp q\sin\varphi)t \csc\varphi} \quad (16)$$

Thus, the ROFrFT of a modulation by $q \in \mathbb{R}$ of a function $e^{\pm jqt} x(t)$ is given by

$$\mathbb{F}_R^\varphi\{\mathbb{M}_q x\}(u_\varphi) = \mathbb{F}_R^\varphi\{e^{\pm jqt} x(t)\}(u_\varphi) = X_R^\varphi(u_\varphi \mp q\sin\varphi) \quad (17)$$

If $\varphi$ equals $\pi/2$ in (17), the expression (17) reduces to $X(\omega - q)$, same as that of FT.

**Theorem 4.3** (**Time-frequency shift property**): *Let $x(t) \in L^1(\mathbb{R})$. Then we get*

$$\mathbb{F}_R^\varphi\{\mathbb{S}_\tau \mathbb{M}_q x\}(u_\varphi) = \mathbb{F}_R^\varphi\{x(t-\tau)\,e^{jqt}\}(u_\varphi) = e^{\frac{j}{2}\tau^2 \cot\varphi - ju_\varphi \tau \csc\varphi + jq\tau}\,X_R^\varphi(u_\varphi - \tau\cos\varphi - q\sin\varphi) \quad (18)$$

*Proof:* The time-frequency shift means replacing $x(t)$ by $x(t-\tau)\,e^{jqt}$ in the integral representation (2) of ROFrFT.

$$\mathbb{F}_R^\varphi\{x(t-\tau)\,e^{jqt}\}(u_\varphi) = \sqrt{1-j\cot\varphi}\int_{-\infty}^{\infty} x(t-\tau)\,e^{jqt}\,e^{\frac{j}{2}t^2 \cot\varphi - j t u_\varphi \csc\varphi}\,dt \quad (19)$$

Simplifying leads to

$$\mathbb{F}_R^\varphi\{x(t-\tau)\,e^{jqt}\}(u_\varphi) = \sqrt{1-j\cot\varphi}\int_{-\infty}^{\infty} x(t-\tau)\,e^{\frac{j}{2}t^2 \cot\varphi - j t (u_\varphi - q\sin\varphi)\csc\varphi}\,dt \quad (20)$$

Now assuming, $(t-\tau) = t'$, i.e., $t = (t'+\tau)$ one solves (20) as

$$\mathbb{F}_R^\varphi\{x(t-\tau)\,e^{jqt}\}(u_\varphi) = \sqrt{1-j\cot\varphi}\int_{-\infty}^{\infty} x(t')\,e^{\frac{j}{2}(t'+\tau)^2 \cot\varphi - j(t'+\tau)(u_\varphi - q\sin\varphi)\csc\varphi}\,dt' \quad (21)$$

$$\mathbb{F}_R^\varphi\{x(t-\tau)\,e^{jqt}\}(u_\varphi) =$$

$$e^{\frac{j}{2}\tau^2 \cot\varphi - ju_\varphi \tau \csc\varphi + jq\tau}\sqrt{1-j\cot\varphi}\int_{-\infty}^{\infty} x(t')\,e^{\frac{j}{2}(t')^2 \cot\varphi - jt'(u_\varphi \csc\varphi - \tau\cot\varphi - q)}\,dt' \quad (22)$$

Simplifying further,

$$\mathbb{F}_R^\varphi\{x(t-\tau)\,e^{jqt}\}(u_\varphi) = e^{\frac{j}{2}\tau^2 \cot\varphi - ju_\varphi \tau \csc\varphi + jq\tau}\sqrt{1-j\cot\varphi}\int_{-\infty}^{\infty} e^{\frac{j}{2}(t')^2 \cot\varphi - jt'(u_\varphi - \tau\cos\varphi - q\sin\varphi)}\,dt'$$

Thus, the ROFrFT of the derivative of a function $x(t-\tau)\,e^{jqt}$ is given by

$$\mathbb{F}_R^\varphi\{\mathbb{S}_\tau \mathbb{M}_q x\}(u_\varphi) = \mathbb{F}_R^\varphi\{x(t-\tau)\,e^{jqt}\}(u_\varphi) = e^{\frac{j}{2}\tau^2 \cot\varphi - ju_\varphi \tau \csc\varphi + jq\tau}\,X_R^\varphi(u_\varphi - \tau\cos\varphi - q\sin\varphi) \quad (23)$$

If $\varphi$ equals $\pi/2$ in (23), the expression (23) reduces to $e^{-j(\omega-q)\tau}\,X(\omega-q)$, same as that of FT.

**Theorem 4.4 (Multiplication by $\cos(\ell t)$):** Let $x(t) \in L^1(\mathbb{R})$. Then we get

$$\mathbb{F}_R^\varphi\{x(t)\cos(\ell t)\}(u_\varphi) = \frac{1}{2}\left[X_R^\varphi(u_\varphi - \ell\sin\varphi) + X_R^\varphi(u_\varphi + \ell\sin\varphi)\right] \quad (24)$$

*Proof:* From (13) above, one have

$$x(t)\,e^{j\ell t} \stackrel{RFrFT}{\longleftrightarrow} X_R^\varphi(u_\varphi - \ell\sin\varphi) \quad (25)$$

$$x(t)\,e^{-j\ell t} \stackrel{RFrFT}{\longleftrightarrow} X_R^\varphi(u_\varphi + \ell\sin\varphi) \quad (26)$$

Therefore, $x(t)\cos(\ell t) \overset{RFrFT}{\longleftrightarrow} \frac{1}{2}\left[X_R^\varphi(u_\varphi - \ell\sin\varphi) + X_R^\varphi(u_\varphi + \ell\sin\varphi)\right]$ (27)

For $\varphi$ equals $\pi/2$ in (27), the expression (27) reduces to $\frac{1}{2}[X(\omega - \ell) + X(\omega + \ell)]$, same as that of FT.

**Theorem 4.5** (**Inversion of Time axis**): *Let $x(t) \in L^1(\mathbb{R})$. Then we get*

$$\mathbb{F}_R^\varphi\{x(-t)\}(u_\varphi) = X_R^\varphi(-u_\varphi) \tag{28}$$

***Proof:*** For deducing the ROFrFT of the time inversion axis, replace $x(t)$ by $x(-t)$ in the integral representation (2) of ROFrFT, one gets

$$\mathbb{F}_R^\varphi\{x(-t)\}(u_\varphi) = \sqrt{1 - j\cot\varphi}\int_{-\infty}^{\infty} x(-t)\, e^{\frac{j}{2}t^2\cot\varphi - j\,t\,u_\varphi\csc\varphi}\,dt \tag{29}$$

Let $-t = t''$ in (29), one obtains

$$\mathbb{F}_R^\varphi\{x(-t)\}(u_\varphi) = -\sqrt{1 - j\cot\varphi}\int_{\infty}^{-\infty} x(t'')\, e^{\frac{j}{2}(t'')^2\cot\varphi + j(t'')u_\varphi\csc\varphi}\,dt'' \tag{30}$$

or $\mathbb{F}_R^\varphi[x(-t)] = \sqrt{1 - j\cot\varphi}\int_{-\infty}^{\infty} x(t'')\, e^{\frac{j}{2}(t'')^2\cot\varphi - j(-u_\varphi)(t'')\csc\varphi}\,dt''$

Thus, $\mathbb{F}_R^\varphi\{x(-t)\}(u_\varphi) = X_R^\varphi(-u_\varphi)$ (31)

If $\varphi$ equals $\pi/2$ in (31), the expression (31) reduces to $X(-\omega)$, same as that of FT.

**Theorem 4.6** (**Multiplication property**): *Let $x(t) \in L^1(\mathbb{R})$. Then we get*

$$\mathbb{F}_R^\varphi\{tx(t)\}(u_\varphi) = (j\sin\varphi)\frac{dX_R^\varphi(u_\varphi)}{du_\varphi} \tag{32}$$

***Proof:*** $X_R^\varphi(u_\varphi) = \sqrt{1 - j\cot\varphi}\int_{-\infty}^{\infty} x(t)\, e^{\frac{j}{2}t^2\cot\varphi - j\,t\,u_\varphi\csc\varphi}\,dt$ (33)

Differentiation of () with respect to $u_\varphi$ gives

$\frac{dX_R^\varphi(u_\varphi)}{du_\varphi} = -(j\csc\varphi)\left\{\sqrt{1 - j\cot\varphi}\int_{-\infty}^{\infty}[tx(t)]\, e^{\frac{j}{2}t^2\cot\varphi - j\,t\,u_\varphi\csc\varphi}\right\}dt$ (34)

The factor in the curly braces of (34) represents the ROFrFT of the function $tx(t)$. Rearranging (34), one obtains

$$T_R^\varphi\{tx(t)\}(u_\varphi) = (j \sin \varphi) \frac{dX_R^\varphi(u_\varphi)}{du_\varphi} \tag{35}$$

The form (35) results in much simpler analytical expression than the conventional expression [17] [19]. If the rotation angle $\varphi$ equals $\pi/2$, the expression (35) reduces to $j\frac{dX(\omega)}{d\omega}$, where $\omega$ is the Fourier transform frequency variable. This shows that the proposed ROFrFT reduces to the conventional FT property.

**Theorem 4.7** (**Differentiation property**): *Let $x(t) \in L^1(\mathbb{R})$. Then we get*

$$T_R^\varphi\left\{\frac{dx(t)}{dt}\right\}(u_\varphi) = \left(ju_\varphi \csc \varphi + \cos \varphi \frac{d}{du_\varphi}\right) X_R^\varphi(u_\varphi) \tag{36}$$

*Proof:* To obtain the ROFrFT of the derivative of a function, replace $x(t)$ by $dx(t)/dt$ in the integral representation (2) and integrate by parts, assuming that $x(t) \to 0$, when $t \to \pm\infty$ [17] (If $x(t)$ is differentiable for all $t$ and vanishes as $t \to \pm\infty$, then the ROFrFT of the derivative of the function can be related to the transform of the undifferentiated function through the use of integration by parts.)

$$T_R^\varphi\left\{\frac{dx(t)}{dt}\right\}(u_\varphi) = \sqrt{1 - j \cot \varphi} \int_{-\infty}^{\infty} \frac{dx(t)}{dt} e^{\frac{j}{2}t^2 \cot \varphi - j\, t\, u_\varphi \csc \varphi}\, dt \tag{37}$$

After simplifying, one obtains

$$T_R^\varphi\left\{\frac{dx(t)}{dt}\right\}(u_\varphi) = -(j \cot \varphi)\, T_R^\varphi[t\, x(t)] + (ju_\varphi \csc \varphi) X_R^\varphi(u_\varphi) \tag{38}$$

$$T_R^\varphi\left\{\frac{dx(t)}{dt}\right\}(u_\varphi) = -(j \cot \varphi)(j \sin \varphi) \frac{dX_R^\varphi(u_\varphi)}{du_\varphi} + (ju_\varphi \csc \varphi) X_R^\varphi(u_\varphi) \tag{39}$$

Thus, the ROFrFT of the derivative of a function $dx(t)/dt$ is given by

$$T_R^\varphi\left\{\frac{dx(t)}{dt}\right\}(u_\varphi) = \left(ju_\varphi \csc \varphi + \cos \varphi \frac{d}{du_\varphi}\right) X_R^\varphi(u_\varphi) \tag{40}$$

So if $\varphi$ equals $\pi/2$ in (40), the expression (40) reduces to $j\omega\, X(\omega)$, same as that of FT.

**Theorem 4.8** (**Mixed product property**): *Let $x(t) \in L^1(\mathbb{R})$. Then we get*

$$\mathbb{F}_R^\varphi \left\{ t \frac{dx(t)}{dt} \right\}(u_\varphi) = \frac{j}{2} \sin(2\varphi) \left[ \frac{d^2 X_R^\varphi(u_\varphi)}{du_\varphi^2} + 2j \csc(2\varphi) u_\varphi \frac{dX_R^\varphi(u_\varphi)}{du_\varphi} + 2j \csc(2\varphi) X_R^\varphi(u_\varphi) \right] \quad (41)$$

***Proof:*** To obtain the ROFrFT of the function $t \frac{dx(t)}{dt}$, use the formulations of Theorems 4.6 and 4.7, which is recapitulated as

$\mathbb{F}_R^\varphi \{tx(t)\}(u_\varphi) = (j \sin \varphi) \frac{dX_R^\varphi(u_\varphi)}{du_\varphi}$ and $\mathbb{F}_R^\varphi \left\{ \frac{dx(t)}{dt} \right\}(u_\varphi) = \left( ju_\varphi \csc \varphi + \cos \varphi \frac{d}{du_\varphi} \right) X_R^\varphi(u_\varphi)$, one easily

obtains

$$\mathbb{F}_R^\varphi \left\{ t \frac{dx(t)}{dt} \right\}(u_\varphi) = j \sin \varphi \frac{d}{du_\varphi} \left\{ ju_\varphi \csc \varphi + \cos \varphi \frac{d}{du_\varphi} \right\} = -u_\varphi \frac{dX_R^\varphi(u_\varphi)}{du_\varphi} - X_R^\varphi(u_\varphi) +$$

$$j \sin \varphi \cos \varphi \frac{d^2 X_R^\varphi(u_\varphi)}{du_\varphi^2} \quad (42)$$

Thus, the ROFrFT of the function $t \frac{dx(t)}{dt}$ is given by

$$\mathbb{F}_R^\varphi \left\{ t \frac{dx(t)}{dt} \right\}(u_\varphi) = \frac{j}{2} \sin(2\varphi) \left[ \frac{d^2 X_R^\varphi(u_\varphi)}{du_\varphi^2} + 2j \csc(2\varphi) u_\varphi \frac{dX_R^\varphi(u_\varphi)}{du_\varphi} + 2j \csc(2\varphi) X_R^\varphi(u_\varphi) \right] \quad (43)$$

These properties are listed below in Table 1:

**Table 1**

**Properties of ROFrFT**

| Operation | Signal $x(t)$ | ROFrFT, $X_R^\varphi(u_\varphi)$ |
|---|---|---|
| **Time shift** | $x(t \pm \tau)$ | $e^{\frac{j}{2}\tau^2 \cot \varphi \pm ju_\varphi \tau \csc \varphi} X_R^\varphi(u_\varphi \pm \tau \cos \varphi)$ |
| **Modulation** | $e^{\pm jqt} x(t)$ | $X_R^\varphi(u_\varphi \mp q \sin \varphi)$ |
| **Time-frequency shift** | $x(t - \tau) e^{jqt}$ | $e^{\frac{j}{2}\tau^2 \cot \varphi - ju_\varphi \tau \csc \varphi + jq\tau} X_R^\varphi(u_\varphi - \tau \cos \varphi - q \sin \varphi)$ |
| **Modulation:** | | |
| **Multiplication by $\cos(bt)$** | $x(t) \cos(bt)$ | |

| Multiplication by $\sin(\ell t)$ | $x(t)\sin(\ell t)$ | $\frac{1}{2}[X_R^\varphi(u_\varphi - \ell\sin\varphi) + X_R^\varphi(u_\varphi + \ell\sin\varphi)]$ |
|---|---|---|
| | | $\frac{1}{2j}[X_R^\varphi(u_\varphi - \ell\sin\varphi) - X_R^\varphi(u_\varphi + \ell\sin\varphi)]$ |
| Multiplication | $t\,x(t)$ | $(j\sin\varphi)\dfrac{dX_R^\varphi(u_\varphi)}{du_\varphi}$ |
| Inversion of time axis | $x(-t)$ | $X_R^\varphi(-u_\varphi)$ |
| Conjugation | $\bar{x}(t)$ | $\bar{X}_R^{-\varphi}(u_\varphi)$ |
| Even function | $\dfrac{x(t)+x(-t)}{2}$ | $\dfrac{X_R^\varphi(u_\varphi)+X_R^\varphi(-u_\varphi)}{2}$ |
| Odd function | $\dfrac{x(t)-x(-t)}{2}$ | $\dfrac{X_R^\varphi(u_\varphi)-X_R^\varphi(-u_\varphi)}{2}$ |
| Differentiation | $dx(t)/dt$ | $\left(ju_\varphi \csc\varphi + \cos\varphi \dfrac{d}{du_\varphi}\right)X_R^\varphi(u_\varphi)$ |
| Mixed product | $t\,dx(t)/dt$ | $\dfrac{j}{2}\sin(2\varphi)\left[\dfrac{d^2 X_R^\varphi(u_\varphi)}{du_\varphi^2} + 2j\csc(2\varphi)\,u_\varphi \dfrac{dX_R^\varphi(u_\varphi)}{du_\varphi} + 2j\csc(2\varphi)\,X_R^\varphi(u_\varphi)\right]$ |

## 5. ROFrFT of simple signals

The following section outlines the detailed analytical derivations of ROFrFTs of some simple signals.

*Delta Signal*:

For $x(t) = \delta(t-\tau)$, the integral representation of ROFrFT (2) is given by

$$X_R^\varphi(u_\varphi) = \sqrt{1-j\cot\varphi}\int_{-\infty}^{\infty}\delta(t-\tau)\,e^{\frac{j}{2}t^2\cot\varphi - j\,t\,u_\varphi\csc\varphi}\,dt$$

Using the delta sifting property [2], one obtains

$$\mathbb{T}_R^\varphi\{\delta(t-\tau)\}(u_\varphi) = X_R^\varphi(u_\varphi) = \sqrt{1-j\cot\varphi}\,\exp\left[\frac{j}{2}\tau^2\cot\varphi - j\tau u_\varphi \csc\varphi\right] \quad (44)$$

***Unit Step Signal:***

For $x(t) = 1$, its integral representation of ROFrFT (2) is given by

$$X_R^\varphi(u_\varphi) = \sqrt{1-j\cot\varphi}\int_{-\infty}^{\infty} e^{\frac{j}{2}t^2\cot\varphi - jtu_\varphi\csc\varphi}\,dt \quad (45)$$

Solving (45) and by knowing of the fact that $\int_{-\infty}^{\infty} e^{j(Ax^2+Bx)}dx = \sqrt{\left(\frac{\pi}{A}\right)}\,e^{j\frac{\pi}{4}}e^{-j\frac{B^2}{4A}}$

Thus, after simplify (45), one obtains

$$\mathbb{T}_R^\varphi\{1\}(u_\varphi) = X_R^\varphi(u_\varphi) = \sqrt{2\pi(1+j\tan\varphi)}\,\exp\left[j\csc(2\varphi)\,u_\varphi^2\right] \quad (46)$$

***Exponential Signal:***

For $x(t) = e^{jqt}$, its integral representation of ROFrFT (2) is given by

$$X_R^\varphi(u_\varphi) = \sqrt{1-j\cot\varphi}\int_{-\infty}^{\infty} e^{jqt}\,e^{\frac{j}{2}t^2\cot\varphi - jtu_\varphi\csc\varphi}\,dt$$

$$X_R^\varphi(u_\varphi) = \sqrt{1-j\cot\varphi}\int_{-\infty}^{\infty} e^{\left(\frac{j}{2}\cot\varphi\right)t^2 - (ju_\varphi\csc\varphi - jq)t}\,dt \quad (47)$$

Solving (47) and by knowing of the fact that $\int_{-\infty}^{\infty} e^{Ax^2-Bx}dx = \sqrt{\left(\frac{\pi}{-A}\right)}\,e^{-\frac{B^2}{4A}}$, one obtains

$$\mathbb{T}_R^\varphi\{e^{jqt}\}(u_\varphi) = \sqrt{2\pi(1+j\tan\varphi)}\,\exp\left[-\frac{j}{2}q^2\tan\varphi - ju_\varphi^2\csc(2\varphi) + jqu_\varphi\sec\varphi\right] \quad (48)$$

***Exponential Signal multiplied by $t$:***

To determine the ROFrFT of $x(t) = te^{jqt}$, make use of Theorem 4.6 (Multiplication property), which is recapitulated below as

$$\mathbb{T}_R^\varphi\{t\,p(t)\}(u_\varphi) = (j\sin\varphi)\,\frac{dP_R^\varphi(u_\varphi)}{du_\varphi},\text{ where the function } p(t) = e^{jqt} \text{ and its ROFrFT } P_R^\varphi(u_\varphi) \text{ is given by}$$

$$\mathbb{T}_R^\varphi\{e^{jqt}\}(u_\varphi) = \sqrt{2\pi(1+j\tan\varphi)}\,\exp\left[-\frac{j}{2}q^2\tan\varphi - ju_\varphi^2\csc(2\varphi) + jqu_\varphi\sec\varphi\right]$$

$$T_R^{\varphi}\{te^{jqt}\}(u_\varphi) = (j \sin \varphi) \frac{d}{du_\varphi}\left\{\sqrt{2\pi(1+j\tan\varphi)} \exp\left[-\frac{j}{2}q^2 \tan\varphi - ju_\varphi^2 \csc(2\varphi) + jqu_\varphi \sec\varphi\right]\right\} \quad (49)$$

Solving (49) and after mathematical manipulations, one obtains

$$T_R^{\varphi}\{te^{jqt}\}(u_\varphi) = \sqrt{2\pi(1+j\tan\varphi)} \, (\csc\varphi - q) \tan\varphi \, \exp\left[-\frac{j}{2}q^2 \tan\varphi - ju_\varphi^2 \csc(2\varphi) + jqu_\varphi \sec\varphi\right]$$

*Linear Chirp Signal*:

For $x(t) = e^{jat^2/2}$, its integral representation of ROFrFT (2) is given by

$$X_R^{\varphi}(u_\varphi) = \sqrt{1-j\cot\varphi} \int_{-\infty}^{\infty} e^{jat^2/2} \, e^{\frac{j}{2}t^2 \cot\varphi - jt u_\varphi \csc\varphi} \, dt$$

$$X_R^{\varphi}(u_\varphi) = \sqrt{1-j\cot\varphi} \int_{-\infty}^{\infty} e^{\left(j\frac{a}{2} + \frac{j}{2}\cot\varphi\right)t^2 - (j u_\varphi \csc\varphi)t} \, dt \quad (50)$$

Solving (50) and by knowing of the fact that $\int_{-\infty}^{\infty} e^{Ax^2 - Bx} dx = \sqrt{\left(\frac{\pi}{-A}\right)} \, e^{-\frac{B^2}{4A}}$, one obtains

$$T_R^{\varphi}\{e^{jat^2/2}\}(u_\varphi) = \sqrt{\frac{2\pi(j+\cot\varphi)}{(a+\cot\varphi)}} \exp\left[-j\frac{\csc(2\varphi)}{(1+a\tan\varphi)} \, u_\varphi^2\right] \quad (51)$$

*Exponential Signal*:

For $x(t) = e^{-t^2/2}$, its integral representation of ROFrFT (2) is given by

$$X_R^{\varphi}(u_\varphi) = \sqrt{1-j\cot\varphi} \int_{-\infty}^{\infty} e^{-t^2/2} \, e^{\frac{j}{2}t^2 \cot\varphi - jt u_\varphi \csc\varphi} \, dt$$

$$X_R^{\varphi}(u_\varphi) = \sqrt{1-j\cot\varphi} \int_{-\infty}^{\infty} e^{-\left(\frac{1}{2} - \frac{j}{2}\cot\varphi\right)t^2 - (j u_\varphi \csc\varphi)t} \, dt \quad (52)$$

Solving (52) and by knowing of the fact that $\int_{-\infty}^{\infty} e^{-Ax^2 - Bx} dx = \sqrt{\left(\frac{\pi}{A}\right)} \, e^{\frac{B^2}{4A}}$, one obtains

$$T_R^{\varphi}\{e^{-t^2/2}\}(u_\varphi) = \sqrt{2\pi} \exp\left[-\frac{1}{2}(1+j\cot\varphi) \, u_\varphi^2\right] \quad (53)$$

*Exponential Signal*:

For $x(t) = e^{-at^2/2}$, its integral representation of ROFrFT (2) is given by

$$X_R^\varphi(u_\varphi) = \sqrt{1-j\cot\varphi}\int_{-\infty}^{\infty} e^{-at^2/2}\ e^{\frac{j}{2}t^2\cot\varphi - jtu_\varphi\csc\varphi}\ dt$$

After simplification, one obtains

$$X_R^\varphi(u_\varphi) = \sqrt{1-j\cot\varphi}\int_{-\infty}^{\infty} e^{-\left(\frac{a}{2}-\frac{j}{2}\cot\varphi\right)t^2 - (ju_\varphi\csc\varphi)t}\ dt \tag{54}$$

Solving (54) and by knowing of the fact that $\int_{-\infty}^{\infty} e^{-Ax^2-Bx}dx = \sqrt{\left(\frac{\pi}{A}\right)}\ e^{\frac{B^2}{4A}}$, one obtains

$$\mathbb{F}_R^\varphi\{e^{-at^2/2}\}(u_\varphi) = \sqrt{\frac{2\pi(1-j\cot\varphi)}{(a-j\cot\varphi)}}\exp\left\{-\frac{1}{2}(a+j\cot\varphi)\left(\frac{\csc^2\varphi}{a^2+\cot^2\varphi}\right)u_\varphi^2\right\} \tag{55}$$

***Exponential Signal with Time Delay:***

For $x(t) = e^{-\frac{a}{2}(t-\tau)^2}$, its integral representation of ROFrFT (2) is given by

$$X_R^\varphi(u_\varphi) = \sqrt{1-j\cot\varphi}\int_{-\infty}^{\infty} e^{-\frac{a}{2}(t-\tau)^2}\ e^{\frac{j}{2}t^2\cot\varphi - jtu_\varphi\csc\varphi}\ dt$$

$$X_R^\varphi(u_\varphi) = \sqrt{1-j\cot\varphi}\int_{-\infty}^{\infty} e^{-\left(\frac{a}{2}-\frac{j}{2}\cot\varphi\right)t^2 - (ju_\varphi\csc\varphi - a\tau)t - \frac{a}{2}\tau^2}\ dt \tag{56}$$

Solving (56) and by knowing of the fact that $\int_{-\infty}^{\infty} e^{-Ax^2-Bx-C}dx = \sqrt{\left(\frac{\pi}{A}\right)}\ e^{\frac{B^2}{4A}-C}$, one obtains

$$\mathbb{F}_R^\varphi\{e^{-\frac{a}{2}(t-\tau)^2}\}(u_\varphi) = \sqrt{\frac{2\pi(1-j\cot\varphi)}{(a-j\cot\varphi)}}\exp\left\{-\frac{(a+j\cot\varphi)}{(a^2+\cot^2\varphi)}\left[\frac{\csc^2\varphi}{2}u_\varphi^2 + ja\tau\csc\varphi\ u_\varphi - \frac{j}{2}a\tau^2\cot\varphi\right]\right\} \tag{57}$$

***Exponential nature Signal:***

For $x(t) = te^{-t^2/2}$, its integral representation of ROFrFT (2) is given by

$$X_R^\varphi(u_\varphi) = \sqrt{1-j\cot\varphi}\int_{-\infty}^{\infty} te^{-t^2/2}\ e^{\frac{j}{2}t^2\cot\varphi - jtu_\varphi\csc\varphi}\ dt$$

$$X_R^\varphi(u_\varphi) = \sqrt{1-j\cot\varphi}\int_{-\infty}^{\infty} te^{-\left(\frac{1}{2}-\frac{j}{2}\cot\varphi\right)t^2 - (ju_\varphi\csc\varphi)t}\ dt \tag{58}$$

Solving (58) and by knowing of the fact that $\int_{-\infty}^{\infty} x\ e^{-Ax^2-Bx}dx = -\sqrt{\left(\frac{\pi}{A}\right)}\frac{B}{A}e^{\frac{B^2}{4A}}$, one obtains

$$\mathbb{F}_R^\varphi\{te^{-t^2/2}\}(u_\varphi) = -j2\sqrt{2\pi}\ (1+j\cot\varphi)\sin\varphi\ \exp\left\{-\frac{1}{2}(1+j\cot\varphi)u_\varphi^2\right\}u_\varphi \tag{59}$$

*Exponential nature Signal with Time Delay*:

For $x(t) = (t-\tau)e^{-(t-\tau)^2/2}$, its integral representation of ROFrFT (2) is given by

$$X_R^\varphi(u_\varphi) = \sqrt{1-j\cot\varphi} \int_{-\infty}^{\infty} (t-\tau)e^{-(t-\tau)^2/2} \; e^{\frac{j}{2}t^2 \cot\varphi - jt u_\varphi \csc\varphi} \, dt \tag{60}$$

Simplifying (60) further, one obtains

$$X_R^\varphi(u_\varphi) = e^{-\frac{\tau^2}{2}}(I_1 - \tau I_2) \tag{61}$$

where the integrals $I_1$ and $I_2$ are defined as

$$I_1 = \sqrt{1-j\cot\varphi} \int_{-\infty}^{\infty} t \exp\left[-\left(\frac{1}{2}-\frac{j}{2}\cot\varphi\right)t^2 - \left(ju_\varphi \csc\varphi - \tau\right)\right] dt \tag{62}$$

and

$$I_2 = \sqrt{1-j\cot\varphi} \int_{-\infty}^{\infty} \exp\left[-\left(\frac{1}{2}-\frac{j}{2}\cot\varphi\right)t^2 - \left(ju_\varphi \csc\varphi - \tau\right)\right] dt \tag{63}$$

Solving (62) and (63), and by knowing of the fact that $\int_{-\infty}^{\infty} x\, e^{-Ax^2-Bx} dx = -\sqrt{\left(\frac{\pi}{A}\right)} \frac{B}{A} e^{\frac{B^2}{4A}}$ and $\int_{-\infty}^{\infty} e^{-Ax^2-Bx} dx = \sqrt{\left(\frac{\pi}{A}\right)} e^{\frac{B^2}{4A}}$, one obtains

$$I_1 = -2\sqrt{2\pi}\left(\frac{ju_\varphi \csc\varphi - \tau}{1-j\cot\varphi}\right) \exp\left[\frac{1}{2}\frac{(ju_\varphi \csc\varphi - \tau)^2}{(1-j\cot\varphi)}\right] \tag{64}$$

and $I_2 = \sqrt{2\pi} \exp\left[\frac{1}{2}\frac{(ju_\varphi \csc\varphi - \tau)^2}{(1-j\cot\varphi)}\right]$ \hfill (65)

From (64) and (65), (61) becomes

$$X_R^\varphi(u_\varphi) = e^{-\frac{\tau^2}{2}}\left\{-2\sqrt{2\pi}\left(\frac{ju_\varphi \csc\varphi - \tau}{1-j\cot\varphi}\right) \exp\left[\frac{1}{2}\frac{(ju_\varphi \csc\varphi - \tau)^2}{(1-j\cot\varphi)}\right] - \tau\sqrt{2\pi} \exp\left[\frac{1}{2}\frac{(ju_\varphi \csc\varphi - \tau)^2}{(1-j\cot\varphi)}\right]\right\} \tag{66}$$

Simplifying further, (66) becomes

$$\mathbb{F}_R^\varphi\{(t-\tau)e^{-(t-\tau)^2/2}\}(u_\varphi) = \sqrt{2\pi}\left(\frac{\tau(1+j\cot\varphi) - j2u_\varphi \csc\varphi}{(1-j\cot\varphi)}\right) \exp\left(-\frac{\tau^2}{2}\right) \exp\left[\frac{1}{2}\frac{(ju_\varphi \csc\varphi - \tau)^2}{(1-j\cot\varphi)}\right] \tag{67}$$

The ROFrFTs of different signals are listed below in Table 2:

**Table 2**

**ROFrFTs of some signals**

| Signal $x(t)$ | ROFrFT with angle $\varphi$, $X_R^\varphi(u_\varphi)$ |
|---|---|
| $\delta(t-\tau)$ | $\sqrt{1-j\cot\varphi}\,\exp\left[\frac{j}{2}\tau^2\cot\varphi - j\tau u_\varphi\csc\varphi\right]$ |
| $1$ | $\sqrt{2\pi(1+j\tan\varphi)}\,\exp[j\csc(2\varphi)\,u_\varphi^2]$ |
| $e^{\pm jqt}$ | $\sqrt{2\pi(1+j\tan\varphi)}\,\exp\left[-\frac{j}{2}q^2\tan\varphi - ju_\varphi^2\csc(2\varphi) \pm jqu_\varphi\sec\varphi\right]$ |
| $te^{\pm jqt}$ | $\sqrt{2\pi(1+j\tan\varphi)}\,(\csc\varphi$ $\mp q)\tan\varphi\,\exp\left[-\frac{j}{2}q^2\tan\varphi - ju_\varphi^2\csc(2\varphi) \pm jqu_\varphi\sec\varphi\right]$ |
| $e^{\pm jat^2/2}$ | $\sqrt{\frac{2\pi(j+\cot\varphi)}{(\pm a+\cot\varphi)}}\,\exp\left[-j\frac{\csc(2\varphi)}{(1\pm a\tan\varphi)}\,u_\varphi^2\right]$ |
| $e^{-t^2/2}$ | $\sqrt{2\pi}\,\exp\left[-\frac{1}{2}(1+j\cot\varphi)\,u_\varphi^2\right]$ |
| $e^{-at^2/2}$ | $\sqrt{\frac{2\pi(1-j\cot\varphi)}{(a-j\cot\varphi)}}\,\exp\left\{-\frac{1}{2}(a+j\cot\varphi)\left(\frac{\csc^2\varphi}{a^2+\cot^2\varphi}\right)u_\varphi^2\right\}$ |
| $e^{-\frac{a}{2}(t-\tau)^2}$ | $\sqrt{\frac{2\pi(1-j\cot\varphi)}{(a-j\cot\varphi)}}\,\exp\left\{-\frac{(a+j\cot\varphi)}{(a^2+\cot^2\varphi)}\left[\frac{\csc^2\varphi}{2}u_\varphi^2 + ja\tau\csc\varphi\,u_\varphi - \frac{j}{2}a\tau^2\cot\varphi\right]\right\}$ |
| $te^{-t^2/2}$ | $-j2\sqrt{2\pi}\,(1+j\cot\varphi)\sin\varphi\,u_\varphi\exp\left\{-\frac{1}{2}(1+j\cot\varphi)\,u_\varphi^2\right\}$ |
| $(t-\tau)e^{-(t-\tau)^2/2}$ | $\sqrt{2\pi}\left(\frac{\tau(1+j\cot\varphi)-j2u_\varphi\csc\varphi}{(1-j\cot\varphi)}\right)\exp\left(-\frac{\tau^2}{2}\right)\exp\left[\frac{1}{2}\frac{(ju_\varphi\csc\varphi-\tau)^2}{(1-j\cot\varphi)}\right]$ |

## 6. Convolution Theorem associated with ROFrFT

**Theorem 6.1** *For any two functions $f, g \in L^1(\mathbb{R})$, let $F_R^\varphi$, $G_R^\varphi$ denote the ROFrFT of $f, g$, respectively. The convolution operator of the ROFrFT is defined as*

$$(f \circledast_R g)(t) = \int_{-\infty}^{\infty} f(\tau)\,g(t-\tau)\,W_{cv}(\tau,t)\,d\tau \tag{68}$$

where, $W_{cv}(\tau, t) = e^{j\tau(\tau-t)\cot\varphi}$. Then, the ROFrFT of the convolution of two complex functions is given by

$$\mathbb{T}_R^\varphi\{f \circledast_R g\}(u_\varphi) = \frac{1}{\sqrt{1-j\cot\varphi}} F_R^\varphi(u_\varphi) G_R^\varphi(u_\varphi) \tag{69}$$

*Proof:* From the definition of ROFrFT 2() and the ROFrFT convolution (68), one obtains

$$\mathbb{T}_S^\varphi\{f \circledast_R g\}(u_\varphi) = \sqrt{1-j\cot\varphi} \int_{-\infty}^{\infty} \{f(t) \circledast_R g(t)\} e^{\frac{j}{2}t^2\cot\varphi - j\, t\, u_\varphi \csc\varphi} dt \tag{70}$$

$$\mathbb{T}_S^\varphi\{f \circledast_R g\}(u_\varphi) = \sqrt{1-j\cot\varphi} \int_{-\infty}^{\infty} \left\{\int_{-\infty}^{\infty} f(\tau) g(t-\tau) W_{cv}(\tau,t)\, d\tau\right\} e^{\frac{j}{2}t^2\cot\varphi - j\, t\, u_\varphi \csc\varphi} dt \tag{71}$$

For solving (71), letting $(t - \tau) = \zeta$

$$\mathbb{T}_S^\varphi\{f \circledast_R g\}(u_\varphi) = \sqrt{1-j\cot\varphi} \int_{-\infty}^{\infty}\int_{-\infty}^{\infty} f(\tau)\, g(\zeta)\, e^{-j\tau\zeta\cot\varphi} e^{\frac{j}{2}(\zeta+\tau)^2\cot\varphi - j(\zeta+\tau)u_\varphi \csc\varphi}\, d\tau\, d\zeta \tag{72}$$

Rearranging and multiplying numerator and denominator of (72) by $\sqrt{1-j\cot\varphi}$, one obtains

$$\mathbb{T}_S^\varphi\{f \circledast_R g\}(u_\varphi) = \sqrt{1-j\cot\varphi} \int_{-\infty}^{\infty} f(\tau)\, e^{\frac{j}{2}\tau^2\cot\varphi - j\, \tau\, u_\varphi \csc\varphi}\, d\tau \,\times$$

$$\sqrt{1-j\cot\varphi} \int_{-\infty}^{\infty} g(\zeta)\, e^{\frac{j}{2}\zeta^2\cot\varphi - j\, \zeta\, u_\varphi \csc\varphi}\, d\zeta \times \frac{1}{\sqrt{1-j\cot\varphi}} \tag{73}$$

By the definition of ROFrFT, the above expression (73) reduces to

$$\mathbb{T}_S^\varphi\{f \circledast_R g\}(u_\varphi) = \frac{1}{\sqrt{1-j\cot\varphi}} F_R^\varphi(u_\varphi) G_R^\varphi(u_\varphi), \tag{74}$$

which proves the theorem in ROFrFT domain.

It is to be noted from (74) that this new proposed definition of FrFT makes the convolution theorem more exciting — the convolution theorem in ROFrFT domain gets reduced in the same form as that of conventional Euclidean Fourier transformation. That is to say, the convolution of two signals of interest in fractional domain gets reduced to simple multiplication of their fractional frequency transforms. This is an added advantage of the proposed new definition of FrFT, which will find widespread applications in fractional filter design in various signal processing systems like communication, radar, sonar, seismic, biomedical processing and to name a few.

*Special case*:

For the Euclidean FT, the rotation angle $\varphi = \pi/2$, then the expression (74) reduces to

$$\mathbb{F}_S^{\pi/2}\{f \circledast_R g\}(u_{\pi/2}) = F_R^{\pi/2}(u_{\pi/2}) G_R^{\pi/2}(u_{\pi/2}) = F_R(\omega) G_R(\omega) \qquad (75)$$

This means that the proposed convolution theorem behaves similar to the Euclidean FT, i.e., the convolution in the time-domain is equivalent to the multiplication in the reduced fractional frequency domain, where $u_{\pi/2} = \omega$.

Some properties associated with the convolution theorem in ROFrFT domain are illustrated below:

***Property 1*** (**Shift convolution**). *Let $f, g \in L^1(\mathbb{R})$. The ROFrFT of $\mathbb{S}_d f \circledast_R g$ and $f \circledast_R \mathbb{S}_d g$ is given by*

$$\mathbb{F}_R^{\varphi}\{\mathbb{S}_d f \circledast_R g\}(u_\varphi) = \frac{1}{\sqrt{1-j\cot\varphi}} e^{-ju_\varphi d \csc\varphi + \frac{j}{2}d^2 \cot\varphi} F_R^{\varphi}(u_\varphi - d\cot\varphi) G_R^{\varphi}(u_\varphi) \qquad (76)$$

$$\mathbb{F}_R^{\varphi}\{f \circledast_R \mathbb{S}_d g\}(u_\varphi) = \frac{1}{\sqrt{1-j\cot\varphi}} e^{-ju_\varphi d \csc\varphi + \frac{j}{2}d^2 \cot\varphi} F_R^{\varphi}(u_\varphi) G_R^{\varphi}(u_\varphi - d\cot\varphi) \qquad (77)$$

where, the symbol $\mathbb{S}_d$ represents the shift operator of a function by delay $d$ i.e., $\mathbb{S}_d x(t) = x(t-d), d \in \mathbb{R}$.

***Proof:*** The shift convolution operator $\mathbb{S}_d f \circledast_R g$ is given by

$$(\mathbb{S}_d f \circledast_R g)(t) = \int_{-\infty}^{\infty} f(\tau - d) g(t - \tau) W_{cv}(\tau, t) d\tau \qquad (78)$$

where, $W_{cv}(\tau, t) = e^{j\tau(\tau-t)\cot\varphi}$. It implies

$$(\mathbb{S}_d f \circledast_R g)(t) = \int_{-\infty}^{\infty} f(\tau - d) g(t - \tau) e^{j\tau(\tau-t)\cot\varphi} d\tau \qquad (79)$$

Now, from the definition of ROFrFT (2), one obtains

$$\mathbb{F}_R^{\varphi}\{\mathbb{S}_d f \circledast_R g\}(u_\varphi) = \sqrt{1-j\cot\varphi} \int_{-\infty}^{\infty} \{\mathbb{S}_d f(t) \circledast_R g(t)\} e^{\frac{j}{2}t^2 \cot\varphi - j t u_\varphi \csc\varphi} dt \qquad (80)$$

Simplifying (80) further, one obtains

$$\mathbb{F}_R^{\varphi}\{\mathbb{S}_d f \circledast_R g\}(u_\varphi) = \sqrt{1-j\cot\varphi} \int_{-\infty}^{\infty} \left\{\int_{-\infty}^{\infty} f(\tau - d) g(t - \tau) e^{j\tau(\tau-t)\cot\varphi} d\tau \right\} e^{\frac{j}{2}t^2 \cot\varphi - j t u_\varphi \csc\varphi} dt$$

Solve above expression by letting $(t - \tau) = p$, one obtains

$$\mathbb{F}_R^\varphi\{\mathbb{S}_d f \circledast_R g\}(u_\varphi) = \sqrt{1 - j \cot \varphi} \int_{-\infty}^{\infty} f(\tau - d) e^{-j u_\varphi \tau \csc \varphi + \frac{j}{2} \tau^2 \cot \varphi} d\tau \times$$

$$\int_{-\infty}^{\infty} g(p) e^{-j u_\varphi p \csc \varphi + \frac{j}{2} p^2 \cot \varphi} dp$$

$$\mathbb{F}_R^\varphi\{\mathbb{S}_d f \circledast_R g\}(u_\varphi) = \int_{-\infty}^{\infty} f(\tau - d) e^{-j u_\varphi \tau \csc \varphi + \frac{j}{2} \tau^2 \cot \varphi} d\tau \times G_R^\varphi(u_\varphi) \qquad (81)$$

Further, by letting $(\tau - d) = z$, and multiplying numerator and denominator of (81) by $\sqrt{1 - j \cot \varphi}$, one solves to get

$$\mathbb{F}_R^\varphi\{\mathbb{S}_d f \circledast_R g\}(u_\varphi) = \sqrt{1 - j \cot \varphi} \; e^{-j u_\varphi d \csc \varphi + \frac{j}{2} d^2 \cot \varphi} \int_{-\infty}^{\infty} f(z) e^{-j(u_\varphi - d \cot \varphi) \csc \varphi z + \frac{j}{2} z^2 \cot \varphi} dz \times$$

$$G_R^\varphi(u_\varphi) \times \frac{1}{\sqrt{1 - j \cot \varphi}}$$

$$\mathbb{F}_R^\varphi\{\mathbb{S}_d f \circledast_R g\}(u_\varphi) = \frac{1}{\sqrt{1 - j \cot \varphi}} e^{-j u_\varphi d \csc \varphi + \frac{j}{2} d^2 \cot \varphi} F_R^\varphi(u_\varphi - d \cot \varphi) G_R^\varphi(u_\varphi),$$

which proves the shift convolution property (76).

Similarly, for solving $\mathbb{F}_R^\varphi\{f \circledast_R \mathbb{S}_d g\}(u_\varphi)$ and utilizing the shift convolution operator of function $f \circledast_R \mathbb{S}_d g$ as $\int_{-\infty}^{\infty} f(\tau) g(t - \tau - d) W_{cv}(\tau, t) d\tau$, where, $W_{cv}(\tau, t) = e^{j \tau (\tau - t) \cot \varphi}$ and based on the previous steps, one obtains

$$\mathbb{F}_R^\varphi\{f \circledast_R \mathbb{S}_d g\}(u_\varphi) = \frac{1}{\sqrt{1 - j \cot \varphi}} e^{-j u_\varphi d \csc \varphi + \frac{j}{2} d^2 \cot \varphi} F_R^\varphi(u_\varphi) G_R^\varphi(u_\varphi - d \cot \varphi), \qquad (82)$$

which proves the shift convolution property (77) in ROFrFT domain.

Thus, (76) and (77) indicates that if we apply a linear time delay to one signal in the time domain and fractional convolve it with the another time domain signal, then the ROFrFT of the convolved signal is identical to the multiplications of the ROFrFTs of the respective signals, except that one of the signal has been shifted in the ROFrFT domain by an amount dependent on the change in time shift in the time domain, and there is a multiplication with the complex harmonic dependent on the time shift.

*Special case*:

For the Euclidean FT, the rotation angle $\varphi = \pi/2$, then the expression (76) and (77) reduces to

$$\mathbb{T}_R^{\pi/2}\{\mathbb{S}_d f \circledast_R g\}(u_{\pi/2}) = e^{-ju_{\pi/2}d} F_R^{\pi/2}(u_{\pi/2}) G_R^{\pi/2}(u_{\pi/2})$$

i.e, $\mathcal{F}\{\mathbb{S}_d f \circledast_R g\}(\omega) = e^{-j\omega d} F_R(\omega) G_R(\omega)$ (83)

$$\mathbb{T}_R^{\pi/2}\{f \circledast_R \mathbb{S}_d g\}(u_{\pi/2}) = e^{-ju_{\pi/2}d} F_R^{\pi/2}(u_{\pi/2}) G_R^{\pi/2}(u_{\pi/2})$$

i.e, $\mathcal{F}\{f \circledast_R \mathbb{S}_d g\}(\omega) = e^{-j\omega d} F_R(\omega) G_R(\omega)$ (84)

This means that the proposed shift convolution property behaves similar to the Euclidean FT, as is evident from (83) and (84), respectively.

*Property 2* (**Modulation convolution**). *Let $f, g \in L^1(\mathbb{R})$. The ROFrFT of $\mathbb{M}_q f \circledast_R g$ and $f \circledast_R \mathbb{M}_q g$ is given by*

$$\mathbb{T}_R^\varphi\{\mathbb{M}_q f \circledast_R g\}(u_\varphi) = \frac{1}{\sqrt{1-j\cot\varphi}} F_R^\varphi(u_\varphi - q\sin\varphi) G_R^\varphi(u_\varphi) \tag{85}$$

$$\mathbb{T}_R^\varphi\{f \circledast_R \mathbb{M}_q g\}(u_\varphi) = \frac{1}{\sqrt{1-j\cot\varphi}} F_R^\varphi(u_\varphi) G_R^\varphi(u_\varphi - q\sin\varphi) \tag{86}$$

where, the symbol $\mathbb{M}_q$ represents the modulation operator, i.e., the modulation by $q$ of a function $x(t)$, $\mathbb{M}_q x(t) = e^{jqt} x(t)$, $q \in \mathbb{R}$.

*Proof:* The modulation convolution operator $\mathbb{M}_q f \circledast_R g$ is given by

$$(\mathbb{M}_q f \circledast_R g)(t) = \int_{-\infty}^{\infty} e^{jq\tau} f(\tau) g(t-\tau) W_{cv}(\tau, t) d\tau \tag{87}$$

where, $W_{cv}(\tau, t) = e^{j\tau(\tau-t)\cot\varphi}$. It implies

$$(\mathbb{M}_q f \circledast_R g)(t) = \int_{-\infty}^{\infty} e^{jq\tau} f(\tau) g(t-\tau) e^{j\tau(\tau-t)\cot\varphi} d\tau \tag{88}$$

Now, from the definition of ROFrFT (2), one obtains

$$\mathbb{F}_R^\varphi\{\mathbb{M}_q f \circledast_R g\}(u_\varphi) = \sqrt{1-j\cot\varphi} \int_{-\infty}^{\infty}\{\mathbb{M}_q f(t) \circledast_R g(t)\} e^{-jt u_\varphi \csc\varphi + \frac{j}{2}t^2 \cot\varphi} dt \tag{89}$$

Simplifying (89) further, one obtains

$$\mathbb{F}_R^\varphi\{\mathbb{M}_q f \circledast_R g\}(u_\varphi) = \sqrt{1-j\cot\varphi} \int_{-\infty}^{\infty}\int_{-\infty}^{\infty} f(\tau) g(t-\tau) \, e^{jq\tau + j\tau(\tau-t)\cot\varphi - ju_\varphi t \csc\varphi + \frac{j}{2}t^2 \cot\varphi} d\tau \, dt$$

(90)

By letting $(t - \tau) = v$, and multiplying numerator and denominator of (90) by $\sqrt{1 - j \cot \varphi}$, (90) reduces to

$$\mathbb{F}_R^\varphi\{\mathbb{M}_q f \circledast_R g\}(u_\varphi) = \sqrt{1-j\cot\varphi} \int_{-\infty}^{\infty} f(\tau) \, e^{-j(u_\varphi - q\sin\varphi)\tau \csc\varphi + \frac{j}{2}\tau^2 \cot\varphi} d\tau \times$$

$$\sqrt{1-j\cot\varphi} \int_{-\infty}^{\infty} h(v) \, e^{-ju_\varphi v \csc\varphi + \frac{j}{2}v^2 \cot\varphi} dv \times \frac{1}{\sqrt{1-j\cot\varphi}}$$

Simplifying further, one obtains

$$\mathbb{F}_R^\varphi\{\mathbb{M}_q f \circledast_R g\}(u_\varphi) = \frac{1}{\sqrt{1-j\cot\varphi}} F_R^\varphi(u_\varphi - q\sin\varphi) \, G_R^\varphi(u_\varphi), \tag{91}$$

which proves the modulation convolution property in ROFrFT domain.

Similarly, for solving $\mathbb{F}_R^\varphi\{f \circledast_R \mathbb{M}_q g\}(u_\varphi)$ and utilizing the modulation convolution operator of function $f \circledast_R \mathbb{M}_q g$ as $\int_{-\infty}^{\infty} f(\tau) \, e^{jq(t-\tau)} g(t-\tau) W_{cv}(\tau,t) \, d\tau$, where, $W_{cv}(\tau,t) = e^{j\tau(\tau-t)\cot\varphi}$ and based on the previous steps, one obtains

$$\mathbb{F}_R^\varphi\{f \circledast_R \mathbb{M}_q g\}(u_\varphi) = \frac{1}{\sqrt{1-j\cot\varphi}} F_R^\varphi(u_\varphi) \, G_R^\varphi(u_\varphi - q\sin\varphi), \tag{92}$$

which proves the modulation convolution property in ROFrFT domain.

Thus, (85) and (86) indicates that if we apply a linear change in phase to one signal in the time domain and fractional convolve it with the another time domain signal, then the ROFrFT of the convolved signal is identical to the multiplications of the ROFrFTs of the respective signals, except that one of the signal has been shifted in the ROFrFT domain by an amount dependent on the change in phase in the time domain, with an amplitude factor dependent on the rotation angle $\varphi$.

*Special case*:

In case of FT, (85) and (86) reduces to (for $\varphi = \pi/2$)

$$\mathbb{T}_R^{\pi/2}\{\mathbb{M}_q f \circledast_R g\}(u_{\pi/2}) = F_R^{\pi/2}(u_{\pi/2} - q) G_R^{\pi/2}(u_{\pi/2}) = F_R(\omega - q) G_R(\omega),$$

i.e., $\mathcal{F}\{\mathbb{M}_q f \circledast_R g\}(\omega) = F_R(\omega - q) G_R(\omega)$ (93)

$$\mathbb{T}_R^{\pi/2}\{f \circledast_R \mathbb{M}_q g\}(u_{\pi/2}) = F_R^{\pi/2}(u_{\pi/2}) G_R^{\pi/2}(u_{\pi/2} - q)$$

$$\mathcal{F}\{f \circledast_R \mathbb{M}_q g\}(\omega) = F_R(\omega) G_R(\omega - q) \tag{94}$$

This means that the proposed modulation convolution property behaves similar to the Euclidean FT, as is evident from (93) and (94), respectively.

*Property 3* (**Time-Frequency shift convolution**). *Let $f, g \in L^1(\mathbb{R})$. The ROFrFT of $\mathbb{M}_q \mathbb{S}_d f \circledast_R g$ and $f \circledast_R \mathbb{M}_q \mathbb{S}_d g$ is given by*

$$\mathbb{T}_R^\varphi\{\mathbb{M}_q \mathbb{S}_d f \circledast_R g\}(u_\varphi) = \frac{1}{\sqrt{1-j\cot\varphi}} e^{-j(u_\varphi - q\sin\varphi)d\csc\varphi + \frac{j}{2}d^2\cot\varphi} F_R^\varphi(u_\varphi - q\sin\varphi - d\cot\varphi) G_R^\varphi(u_\varphi)$$

(95)

$$\mathbb{T}_R^\varphi\{f \circledast_R \mathbb{M}_q \mathbb{S}_d g\}(u_\varphi) = \frac{1}{\sqrt{1-j\cot\varphi}} e^{-j(u_\varphi - q\sin\varphi)d\csc\varphi + \frac{j}{2}d^2\cot\varphi} F_R^\varphi(u_\varphi) G_R^\varphi(u_\varphi - q\sin\varphi - d\cot\varphi)$$

(96)

where, the symbol $\mathbb{S}_d$ and $\mathbb{M}_q$ represents the shift operator of a function by delay $d$ and the modulation operator of a function by $q$, i.e., for the function $x(t)$, $\mathbb{S}_d x(t) = x(t-d)$, $d \in \mathbb{R}$ and $\mathbb{M}_q x(t) = e^{jqt} x(t)$, $q \in \mathbb{R}$.

*Proof:* The time-frequency shift convolution operator is given by

$$(\mathbb{M}_q \mathbb{S}_d f \circledast_R g)(t) = \int_{-\infty}^{\infty} e^{jq\tau} f(\tau - d) g(t - \tau) W_{cv}(\tau, t) d\tau \tag{97}$$

where, $W_{cv}(\tau, t) = e^{j\tau(\tau-t)\cot\varphi}$. It implies

$$\left(\mathbb{M}_q \mathbb{S}_d f \circledast_R g\right)(t) = \int_{-\infty}^{\infty} e^{jq\tau} f(\tau - d)\, g(t - \tau)\, e^{j\tau(\tau - t)\cot\varphi}\, d\tau \tag{98}$$

The ROFrFT of (97) is obtained as

$$\mathbb{F}_R^{\varphi}\{\mathbb{M}_q \mathbb{S}_d f \circledast_R g\}(u_\varphi) = \sqrt{1 - j\cot\varphi} \int_{-\infty}^{\infty} \{\mathbb{M}_q \mathbb{S}_d f(t) \circledast_R g(t)\} e^{-j t u_\varphi \csc\varphi + \frac{j}{2} t^2 \cot\varphi} dt \tag{99}$$

Simplifying (99) further, one obtains

$$\mathbb{F}_R^{\varphi}\{\mathbb{M}_q \mathbb{S}_d f \circledast_R g\}(u_\varphi)$$

$$= \sqrt{1 - j\cot\varphi} \int_{-\infty}^{\infty}\int_{-\infty}^{\infty} f(\tau - d)\, g(t - \tau)\, e^{jq\tau + j\tau(\tau - t)\cot\varphi - j t u_\varphi \csc\varphi + \frac{j}{2} t^2 \cot\varphi}\, d\tau\, dt$$

$$\tag{100}$$

By letting $(t - \tau) = \varsigma$, (100) is simplified as

$$\mathbb{F}_R^{\varphi}\{\mathbb{M}_q \mathbb{S}_d f \circledast_R g\}(u_\varphi) = \int_{-\infty}^{\infty} f(\tau - d)\, e^{jq\tau - ju_\varphi \tau \csc\varphi + \frac{j}{2}\tau^2 \cot\varphi}\, d\tau \times$$

$$\sqrt{1 - j\cot\varphi} \int_{-\infty}^{\infty} g(\varsigma)\, e^{-ju_\varphi \varsigma \csc\varphi + \frac{j}{2}\varsigma^2 \cot\varphi}\, d\varsigma \tag{101}$$

Let $(\tau - d) = \xi$, and multiplying numerator and denominator of (101) by $\sqrt{1 - j\cot\varphi}$, (101) reduces to

$$\mathbb{F}_R^{\varphi}\{\mathbb{M}_q \mathbb{S}_d f \circledast_R g\}(u_\varphi) = \frac{1}{\sqrt{1 - j\cot\varphi}} \int_{-\infty}^{\infty} f(\xi)\, e^{-j\xi(u_\varphi - q\sin\varphi - d\cos\varphi)\csc\varphi + \frac{j}{2}\xi^2 \cot\varphi}\, d\xi \times \sqrt{1 - j\cot\varphi} \times$$

$$e^{-ju_\varphi d \csc\varphi + jqd + \frac{j}{2}d^2 \cot\varphi} \times G_R^{\varphi}(u_\varphi) \tag{102}$$

Thus,

$$\mathbb{F}_R^{\varphi}\{\mathbb{M}_q \mathbb{S}_d f \circledast_R g\}(u_\varphi) = \frac{1}{\sqrt{1 - j\cot\varphi}}\, e^{-j(u_\varphi - q\sin\varphi)d\csc\varphi + \frac{j}{2}d^2\cot\varphi}\, F_R^{\varphi}(u_\varphi - q\sin\varphi - d\cot\varphi)\, G_R^{\varphi}(u_\varphi), \tag{103}$$

which proves the time-frequency shift convolution property in ROFrFT domain.

Similarly, for solving $\mathbb{F}_R^\varphi\{f \circledast_R \mathbb{M}_q \mathbb{S}_d g\}(u_\varphi)$ and utilizing the shift and modulation convolution operator of function $f \circledast_R \mathbb{M}_q \mathbb{S}_d g$ as $\int_{-\infty}^{\infty} f(\tau) e^{jq(t-\tau)} g(t - \tau - d) W_{cv}(\tau, t) d\tau$, where, $W_{cv}(\tau, t) = e^{j\tau(\tau-t)\cot\varphi}$ and based on the previous steps, one obtains

$$\mathbb{F}_R^\varphi\{f \circledast_R \mathbb{M}_q \mathbb{S}_d g\}(u_\varphi) = \frac{1}{\sqrt{1-j\cot\varphi}} e^{-j(u_\varphi - q\sin\varphi)d\csc\varphi + \frac{j}{2}d^2\cot\varphi} F_R^\varphi(u_\varphi) G_R^\varphi(u_\varphi - q\sin\varphi - d\cot\varphi),$$

(104)

which proves the time-frequency shift convolution property in ROFrFT domain.

***Special case*:**

In case of FT, (95) and (96) reduces to (for $\varphi = \pi/2$)

$$\mathbb{F}_R^{\pi/2}\{\mathbb{M}_q \mathbb{S}_d f \circledast_R g\}(u_{\pi/2}) = e^{-j(u_{\pi/2}-q)d} F_R^{\pi/2}(u_{\pi/2} - q) G_R^{\pi/2}(u_{\pi/2}),$$

i.e., $\mathcal{F}\{\mathbb{M}_q \mathbb{S}_d f \circledast_R g\}(\omega) = e^{-j(\omega-q)d} F_R(\omega - q) G_R(\omega)$ (105)

$$\mathbb{F}_R^{\pi/2}\{f \circledast_R \mathbb{M}_q \mathbb{S}_d g\}(u_{\pi/2}) = e^{-j(u_{\pi/2}-q)d} F_R^{\pi/2}(u_{\pi/2}) G_R^{\pi/2}(u_{\pi/2} - q)$$

i.e, $\mathcal{F}\{f \circledast_R \mathbb{M}_q \mathbb{S}_d g\}(\omega) = e^{-j(\omega-q)d} F_R(\omega) G_R(\omega - q)$ (106)

This means that the proposed time-frequency shift convolution property behaves similar to the Euclidean FT, as is evident from (105) and (106), respectively.

## 7. Conclusions and Future Scope of Work

In this paper, a new definition of the fractional Fourier transform is introduced, which is considered to be a reduced form of the conventional FrFT — *Reduced Order Fractional Fourier Transform* (ROFrFT). The mathematical definition along with its various properties is derived and it is shown analytically that its definition gets reduced to the conventional Fourier transform definition for the rotational angle of $\pi/2$ radians. Also, different analytical derivation of various kinds of signals is derived, which is much simpler in mathematical form as compared to the conventional FrFT definition. Finally, the new definition of the convolution theorem is presented with its various important properties such as shift, modulation and time-frequency shift convolution properties. It is to be noted that the proposed new definition of FrFT very nicely generalizes the convolution theorem same as that of conventional Fourier transformation, which was not possible with the earlier definitions of conventional fractional Fourier transformation. Thus, new proposed definition of FrFT could prove to be much beneficial for the design of fractional filters or for fractional correlation and also will have an added advantage of simplicity in digital computation, optical implementation and radar/sonar system implementation. In near future, this new definition of FrFT will have great potential to substitute the conventional definition of FrFT in many applications.